\documentclass[emulateapj,epsfig]{article} 
\usepackage[onecolumn]{emulateapj}
\usepackage[]{epsfig}
\hoffset -1.5truecm \lefthead{Menci et al.} \righthead{Bimodal Color Distribution} 
\righthead{} 
\begin{document}
\title{BIMODAL COLOR DISTRIBUTION IN HIERARCHICAL GALAXY FORMATION} \author{N. 
Menci, A. Fontana, E. Giallongo, S. Salimbeni} 
\affil{INAF - Osservatorio Astronomico di Roma, via di Frascati 33, 00040 Monteporzio, Italy}
\smallskip 
\begin{abstract} 
We show how the observed bimodality in the 
color distribution of  galaxies can be  explained in   the  framework  of  the 
hierarchical  clustering picture in terms of the  interplay  between the 
properties of the merging histories and the  feedback/star-formation processes 
in the progenitors of local  galaxies. Using a   semi-analytic  model   of 
hierarchical  galaxy  formation,   we compute the  color distributions  of 
galaxies  with different  luminosities and compare them  with the observations. 
Our fiducial model matches the fundamental properties  of  the observed 
distributions,  namely: 1)  the distribution  of objects  brighter than 
$M_r\lesssim -18$ is clearly bimodal, with a fraction of red objects  increasing 
with  luminosity;  2) for  objects  brighter than $M_r\lesssim  -21$ the  color 
distribution is dominated by red objects with color $u-r\approx 2.2-2.4$; 3) the 
spread on the  distribution of the  red population is  smaller than that  of the 
blue  population;  4)  the  fraction of  red  galaxies   is  larger  in  denser 
environments,  even  for  low-luminosity  objects;  5)  the  bimodality  in 
the distribution persists up to $z\approx 1.5$. We discuss the role of the 
different physical processes included in the model in producing the above 
results. 
\end{abstract}
\section{INTRODUCTION}
Recent observations indicate that the color function of galaxies at low redshift 
is characterized by a bimodal distribution (Strateva et al. 2001; Baldry et  al. 
2004), which   defines two  classes of  galaxies; a  red population,  largely 
contributed by non-star forming galaxies which have formed most of their stellar mass 
at high  redshifts $z$,  and a  blue population  consisting of galaxies actively 
forming stars. The  fraction of galaxies  belonging to the  red population grows 
with the luminosity $L$, being smaller than the blue fraction for  $M_r>  -19.5$ 
and   dominating  the   distribution   for  $M_r>   -21$. There is also a clear 
dependence on the galaxy environment, since red galaxies  typically populate 
the  overdense  regions  (Balogh et  al.  2004).   The  bimodal galaxy 
distribution is present at least up to $z\approx 1-2$ (Bell et al. 2004, 
Giallongo et al. 2005).

Such a sharp, distinctive feature of the galactic population must constitute  a 
powerful constraint  to the   galaxy  formation  models aiming  at deriving  the 
galactic   properties ''ab   initio''. These connect the formation of stars in 
progenitor dark matter (DM) haloes (formed from the collapse of overdense 
regions in the primordial density field) with the assemblage of galaxies through 
repeated merging events involving the DM haloes. 

Such semi-analytic models (SAMs), introduced  by   Kauffman,  White \& 
Guiderdoni (1993) and  Cole et al.  (1994), have been largely  improved (Cole et 
al. 2000; Somerville,  \& Primack  1999)  also through  the  inclusion of 
starbursts  driven  by  merging (Somerville, Primack \& Faber  2001;  Menci  et 
al. 2002)  and by fly-by  events (Menci et al. 2004) between satellite 
galaxies (see also Okamoto \& Nagashima 2003). Recently, hydrodynamical 
simulations have also been able to provide independent insight on galaxy 
evolution on cosmological scale  (Nagamine et al. 2004, 2005; Dave'  et al 
2004).  Although differences exist  among these models, they are now in overall 
agreement not only with the local properties of galaxies but also with their 
statistical properties at high redshifts  both in  the K-band (probing the total 
stellar mass assembled in galaxies) and in the B-  and UV- bands (probing the 
star formation rate). 
According to these models, the tendency  of massive galaxies to be red  and old, 
and of the low mass ones to  be bluer and star-forming, is a natural  outcome of 
the  hierarchical  scenarios.  Indeed, the  progenitors  of present-day massive 
galaxies formed in biased, high-density regions that collapsed at higher $z$ and 
contained  denser  gas. Conversely,  feedback  effects are  mostly  effective at 
prolonging the active phase of star formation at low masses (typically below 
$\sim 10^{11}-10^{12}\,M_{\odot}$).

However, the existence of a clear bimodality in the galaxy color distribution 
constitutes a finer probe for the physical processes included in the models, 
since it suggests a dychotomy in the star--formation histories of galaxies that 
should be originated by deep, fundamental properties of the galaxy formation 
process.

Here we investigate in detail whether the above qualitative trend of the 
hierarchical models may provide a bimodal color distribution in {\it 
quantitative} agreement with the observations. To this aim, we derive in Sect. 2 
the color distribution of galaxies as a function of their luminosity from a 
Monte Carlo version of the SAM model of Menci et al. (2004), and compare in 
Sect.3 the results with the observed distributions.  In Sect. 4 we discuss the 
physical origin of the bimodal distribution in hierarchical scenarios, as it 
emerge from our results. Sect. 5 is devoted to summary and conclusions. 

\section{THE MODEL}

As a baseline, we adopt the SAM   described in Menci et  al. (2002, 2004);  this 
connects the   baryonic processes   (gas cooling,   star formation,   Supernovae 
feedback)  to the  merging  histories  of the  DM  haloes  and of  the  galactic 
sub-haloes there   contained  following the canonical  recipes adopted by  SAMs. 
Starbursts  triggered by galaxy interactions  (fly-by   and   merging    events) 
in common  DM  haloes  are  also included as  described in Menci et  al. (2004). 
However, several changes have been made  in both the DM and the  baryonic sector 
of our model with respect to our previous papers. We describe such changes in 
turn. 

\subsection{The Dark Matter Merging Trees}

To derive the whole color distribution  we run a Monte Carlo version of  our 
SAM, where many  realizations  of  the merging   histories of  present-day DM 
haloes (with different final  masses)  are drawn   by extracting  merging 
probabilities according to  the Extended Press \& Schechter  Theory (Bond et al. 
1991;  Lacey \& Cole   1993). To  generate the  merging trees  we adopt  the 
same  algorithm described in  Cole et   al. (2000),  with a  grid of 25 final 
DM     masses ranging              from              $3.16\cdot 10^9\,M_{\odot}$ 
to $1.8\cdot10^{15}\,M_{\odot}$ with equally spaced logarithmic values, and a 
mass resolution of $5\cdot 10^7\,M_{\odot}$; a total  of 100 realizations are 
drawn for each  final  DM mass.  The evolution of DM haloes described above is 
connected to the processes involving the baryons as described below. 

\subsection{The Radiative Cooling} 
The  mass $m_c$   of cooled  gas in a 
disk  with radius  $r_d$ and  rotation velocity  $v_d$  is derived (Menci   et 
al.  2002,  and Somerville \& Primack 1999) by  computing, for each time-step, 
the increment in the cooling radius  $\Delta r_c$  of the central galaxy in all 
the DM haloes. The  corresponding increment of mass of the cooled gas is 
$\Delta m_c=4\, \pi\,r_c^2\rho_g(r_c)\,\Delta  r_c$, assuming a  simple isothermal gas 
density profile $\rho_g(r)\propto r^{-2}$. The normalization is set as to 
recover the hot gas mass when the density profile is integrated  over the 
volume. After  merging events we determine  whether the  mass of  the largest 
progenitor $m_1$ comprises less than a  fraction $1/2$ of the post-merger mass 
(as in Somerville  \&  Primack  1999);  if  this is the case (i.e., the merging 
partners have comparable masses)  we reset the cooling time  and cooling radius 
to zero (as suggested by recent, aimed hydrodynamical simulation of major galaxy 
merging, see   Cox et al.  2004), and the  gas is reheated  to the virial 
temperature of the  DM  halo. The introduction of  such  a  process suppresses 
the cooling in the massive haloes, which undergo a larger number of such major 
merging events compared to low-mass haloes. 
In addition, the steepness of the faint-end slope of the luminosity function 
(LF hereafter, see Sect. 2.5 and fig. 1) may contribute to account 
for the nonexistence of extremely bright (''monster'') galaxies (see Benson et al. 2003). 
See also Sect. 4 for a discussion on this point. 

\subsection{The Star Formation Law and the Supernovae Feedback}

As for the star formation, we assume the canonical Schmidt form  $\dot 
m_*=m_c/(q\,\tau_d)$, where $\tau_d\equiv r_d/v_d$ and $q$ is  left as  a  free 
parameter.  At each time step, the  mass $\Delta   m_h$ returned from the cold 
gas content of the disk to the hot gas phase due to Supernovae (SNe) activity is 
estimated from canonical energy balance arguments (Kauffman 1996, Kauffmann \& 
Charlot 1998; see also Dekel \& Birnboin 2005) as $\Delta 
m_h=E_{SN}\,\epsilon_0\,\eta_0\,\Delta    m_*/v_c^2$ where $\Delta m_*$ is the 
mass of stars formed in the timestep, $\eta\approx 3-5\cdot 10^{-3}/M_{\odot}$ 
is the number of   SNe  per  unit   solar   mass  (depending on  the assumed 
IMF), $E_{SN}=10^{51}\,{\rm ergs}$ is the energy of ejecta of each SN, and $v_c$ 
is the circular velocity of the galactic halo; $\epsilon_0=0.01-0.5$ is the 
efficiency  of  the  energy  transfer to the cold interstellar gas. The 
above mass $\Delta m_h$ is made available for cooling at the next timestep. The 
model free  parameters   $q=30$  and  $\epsilon_0=0.1$  are chosen as to   match 
the  local B-band LF and  the Tully-Fisher relation  adopting a  Salpeter  IMF.  

\subsection{The Properties of the Disks}

The above   star  formation  law  depends strongly  on   $r_d$  entering the 
timescale $\tau_d$.    The   disk   radius $r_d=r_v\,g(v_c)$  is related  to the 
DM virial radius  $r_v$ by  the function  $g(v_c)$ accounting  for  the 
properties  of  the disk;  we  shall adopt   as our  fiducial choice   the 
model  by  Mo,   Mao  \& White           (1998)            which 
yields            $g(v_c)\approx (\lambda/\sqrt{2})[j_d/(m_c/m)]f_c(v_c)^{-
1/2}\,f_R(v)$, where we take for the  DM spin parameter the average value 
$\lambda=0.05$,  while the ratio of the gas  to DM angular momentum is kept  to 
$j_d=0.05$  as in Mo,  Mao \& White (1998).  The slowly varying functions $f_c$ 
and  $f_R$  account for the gas  density  profile and for the self-gravity 
of the disk; the former is assumed to have the form given by Navarro, Freenk \& 
White (1996); the concentration parameter entering their profile depends on the 
mass, and is computed following the procedure given in the appendix of the above 
paper.

Note that our form of $g(v)$ yields a star formation timescale 
$\tau_d\propto m/m_c$, and hence appreciably affects the star formation in the 
most massive haloe where $\dot m_*$ is not effectively balanced by Supernovae 
feedback. In particular, such haloes are caracterized by a larger $\dot m_*$ at 
high redshifts (where rapid cooling yields larger $m_c/m$ ratios), which implies 
faster consumption of cold gas; at low-$z$, conversely, the longer star 
formation timescales due to the low $m_c/m$ ratio inhibits the formation of 
stars in galaxies within massive DM haloes. 

\subsection{Testing the Model} 

We summarize in fig. 1 some results of the model obtained for our fiducial 
choice of cosmological parameters: $\Omega_0=0.3$, $\Omega_{\Lambda}=0.7$, 
$\Omega_b=0.05$ and $H_0=70\,{\rm km\,s^{-1}\,Mpc^{-1}}$; the metallicity and 
the dust extinction are computed as in Menci et al. (2002), as well as the 
evolution   of the stellar  populations, with  emission derived from synthetic 
spectral energy distributions (Bruzual  \& Charlot  1993), adopting a Salpeter 
IMF. 

To  test  the  consistency  of  our  set  of  cooling,  star formation and 
feedback laws with the available observations, we first compare with the local 
observed distribution of cold gas, stars and of  disk sizes measured  at low-$z$ 
(panels  a, b, c), and with the Tully-Fisher  relation in  the I-band shown  in 
panel d). Then we compare with the B-band (at low $z$) and the UV (at  high $z$) 
luminosity functions (panels e and f),  to  show  that  the instantaneous  star 
formation implemented  in the  model is  consistent with observations over a 
wide range of cosmic times. 

\vspace{0.1cm} 
\begin{center} 
\scalebox{0.55}[0.55]{\rotatebox{0}{\includegraphics{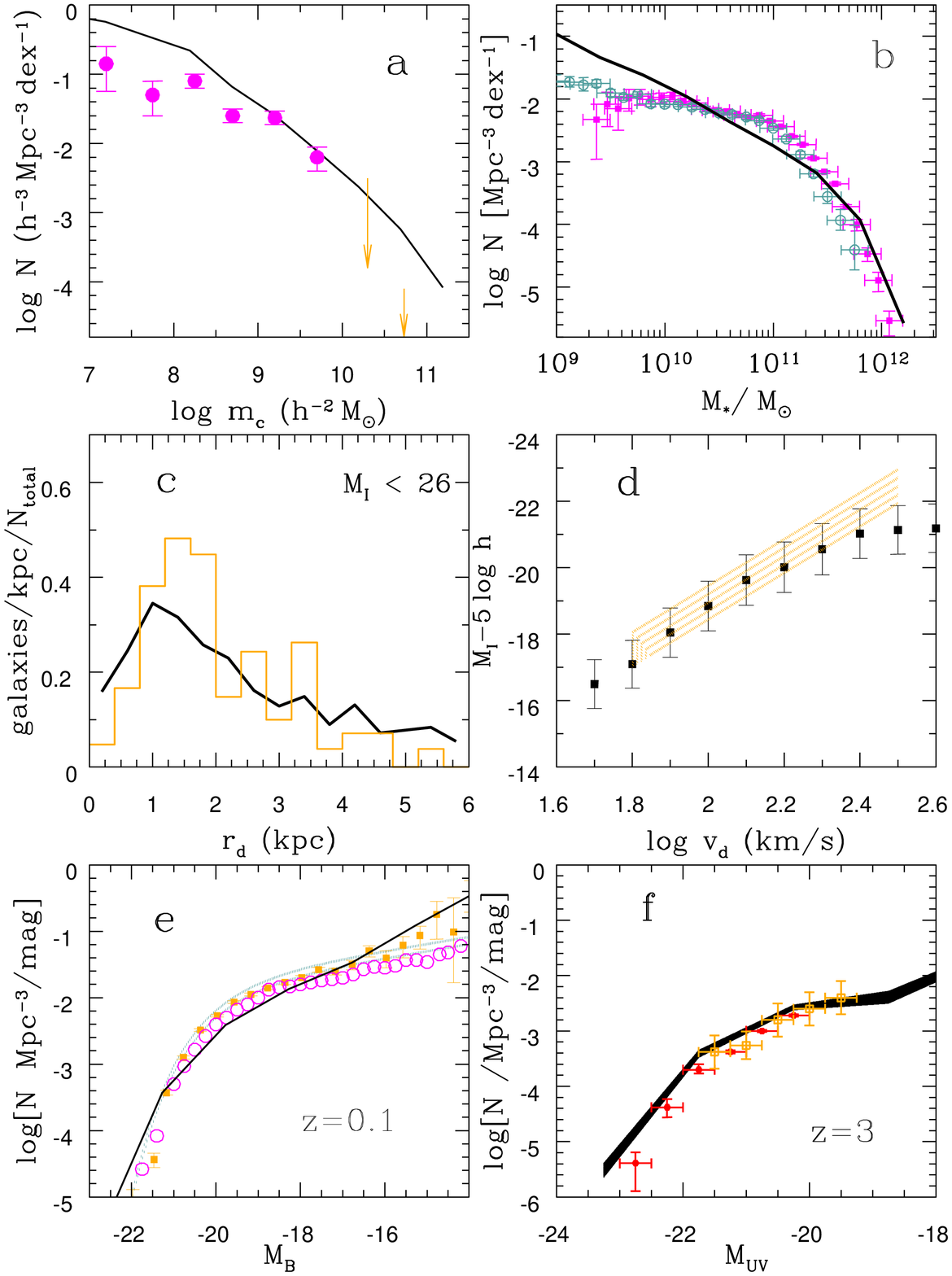}}} 
\end{center} 
{\footnotesize Fig.  1. - Panel a: The mass distribution of 
cold gas is compared with the observed $H_I$ mass function, taken from Zwaan et 
al. (1997). The arrows show upper limits from the Arecibo surveys (see Schneider 
1997). Note that the observed mass in $H_I$ constitutes a lower limit for $m_c$, 
since a part of the cold gas is not in the form of $H_I$. 
\newline
Panel b: We compare the local stellar mass distribution with the data from the 
2dF survey (Cole et al. 2001, filled squares) and from the 2MASS survey (Bell et 
al. 2003, open circles). 
\newline
Panel c: The distribution of disk sizes (exponential scalelengths) from our 
model is compared with the data obtained by Giallongo et al. (2000, histogram) 
for $M_I\leq 26$ galaxies in the Hubble and the NTT Deep Fields at $z=0.4-0.7$.
\newline
Panel d: The Tully-Fisher relation from our model (squares) with the $1\sigma$ 
variance (errorbar); the shaded region represent the region of the $M_I-v$ plane 
spanned by the observations (Mathewson et al. 1992; Willick et al. 1996; 
Giovanelli et al. 1997).
\newline
Panel e: The local B-band luminosity function from our model (solid line), 
assuming a Galaxy dust extiction curve. The shaded area corresponds to the LF 
measured by the Sloan Digital Sky Survey (Blanton et al. 2000), and the circles 
to the data from the 2dFGRS survey (Madgwick et al. 2002). We also show as solid 
squares the data from Zucca et al. (1997). 
\newline
Panel f: The UV luminosity function from our model: the shaded area represent 
the uncertainty in the model results due to assuming different extinction 
curves (Small Magellanic Cloud, Galaxy and Calzetti 1997).  The spectroscopic 
(solid squares) and the photometric (open squares) data are from Steidel et al. 
(1999). 
\vspace{0.1cm}}

Note that the luminosity functions tend to overestimate the number of faint 
galaxies, a long-standing problem of hierarchical models; although this could be 
alleviated by increasing the $v$-dependence of the Supernovae feedback, we 
prefer to keep the scaling of $\Delta m_h$ to the relation $\Delta m_h\sim 
v_c^{-2}$ derived from the energy balance argument (see Sect. 2.3) in order to 
keep the number of free parameters as low as possible; in addition, increasing 
the exponent of the above relation would worsen the agreement with the Tully-
Fisher relation. However, we will focus on the color distribution of relatively 
bright galaxies where the predicted and the observed luminosity function are 
still in good agreement. 

The agreement of our model with the stellar mass functions up  to $z\approx 1.5$ 
and with the K-band luminosity functions and counts up to  the  $z\approx 2$ is 
unchanged with respect to Menci et al. (2004). 

\section{RESULTS}

\subsection{Bimodal Color Distribution at Low and High Redshift}

Here we present the color distributions of  galaxies resulting from our model, 
and compare them with the most distinctive observational results; these concern 
the dependence of the local color distributions on the galaxy luminosity and on 
the environment, and the persistence of the bimodal shape at higher $z$. 

As for the former, we compare in fig. 2 the  color distribution resulting from 
our model for different luminosity bins with the Gaussian  fit to the 
observational points from the sloan survey (SDSS), given in Baldry  et al. 
(2004) for the $u-r$  colors; details on the SDSS $u$  and $r$ bands are given 
by the above authors. We  show our results for three  kinds of dust extiction 
laws (SMC, Galaxy, and  Calzetti 1997); the corresponding color distributions 
are almost identical. 

\vspace{0.1cm} 
\begin{center} 
\scalebox{0.65}[0.65]{\rotatebox{0}{\includegraphics{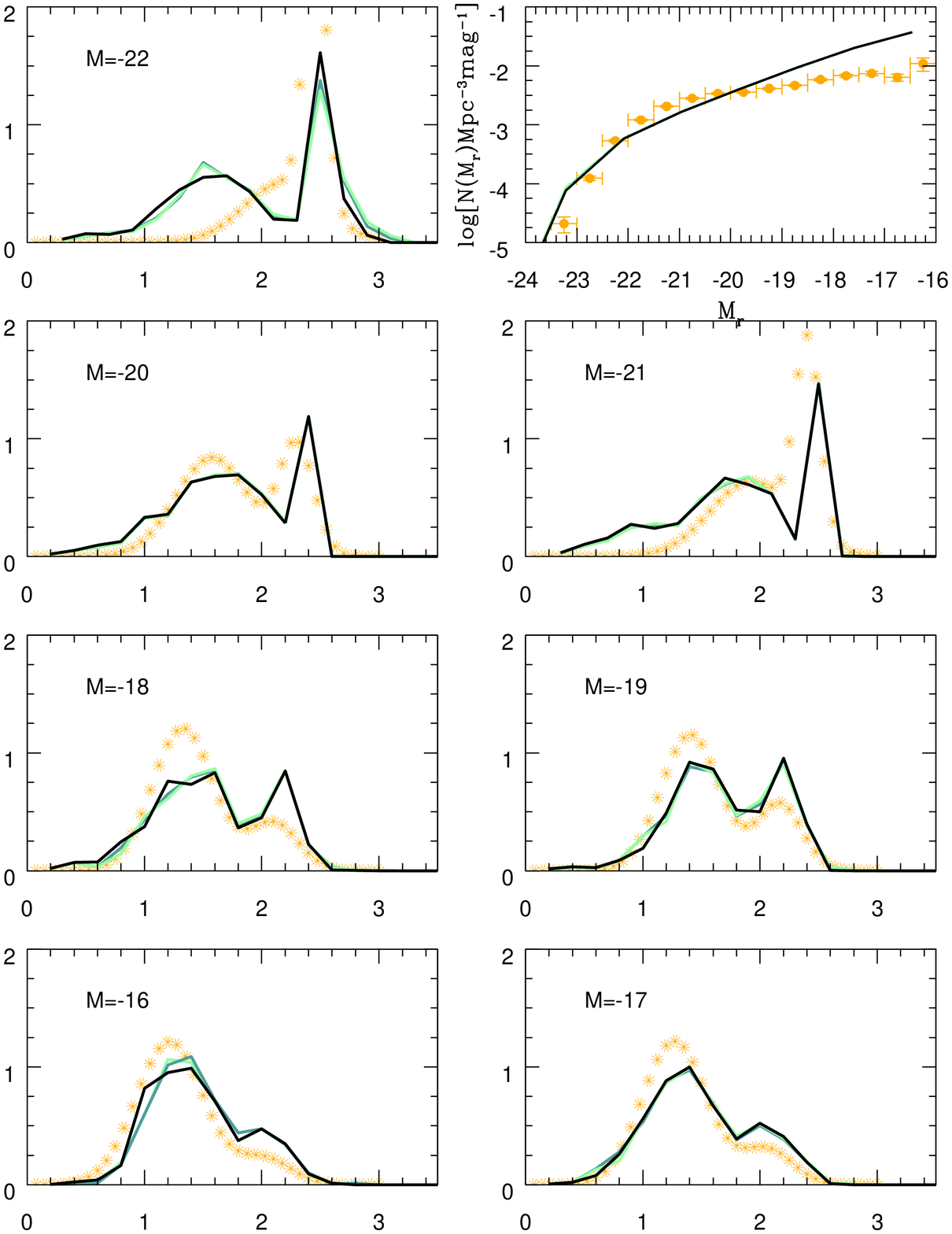}}} 
\end{center} 
\vspace{-0.15cm}
{\footnotesize Fig.  2. -  
Predicted rest-frame $u-r$ 
color distributions (heavy lines) for different  dust extiction laws are 
compared with the  Gaussian fit to  the SDSS data  (from Baldry et al. 2004, 
stars) for different magnitude bins. The distributions are normalized to the 
total number of galaxies in the magnitude bin; the  normalization of the data 
and of the model predictions are given by the r-band LFs shown in the top-right 
panel. The typical size of errorbars is shown in fig. 6.
\vspace{0.1cm}}

The plot shows that hierarchical  galaxy formation may indeed produce  a bimodal 
color distribution, with  the fraction of  red galaxies growing  with increasing 
luminosities as observed. The color ($u-r\approx 2$)  which marks the transition 
between    the two  populations  is   in  good  agreement  with   the   observed 
distributions. In addition, the  model naturally  yields the  correct  behaviour 
of  the  spread,  the  latter  being  $\approx  3$  times  larger  for  the blue 
population. Note however that the  model predicts the  existence of   a residual 
fraction of   blue bright ($M_r\lesssim   -22$) objects. Such  an excess  could 
be alleviated   by including  the feedback from Active Galactic Nuclei (AGNs) in 
the  SAMs;   indeed,  in    our  model,  no  feedback  of  this  kind   has been 
implemented. This shows  that, although AGNs  might  appreciably contribute   to 
suppress  star formation  in part   of the massive  galaxy population,  they are 
not the  origin of  the observed   bimodality in  the color  distribution, whose 
essential features are captured by our SAM. 

We checked the robustness of our predictions against changes in the free 
parameters of our model, within the limits allowed by the simultaneous matching 
of the Tully-Fisher relation and of the B- and UV -band luminosity functions. 
We find that the main features of our models remain stable against most of these 
variations, although the quantitative details are sensitive to the parameter 
choice. For instance, increasing the feedback paremeter $\epsilon_0$ leads to a 
reddening of $\approx 0.3$ of the transition color that marks the partition 
between the red and the blue population. Increasing the star formation timescale 
within the limit $q\lesssim 60$ results into a progressively less pronounced 
bimodality and in more concentrated distributions around the same average 
colour.  We have tested that changing the normalization of the dust 
absorption or of the stellar yield (and hence of metallicity, implemeted as 
described in Menci et al. 2002) does not affect appreciabely our color 
distributions if such normalizations are chosen as to match the B- and UV- band 
luminosity functions from $z=0$ to $z=3$. The only significant exception to 
this rule is the critical dependence on the prescriptions for the star--
formation laws: we shall come back on this issue later in the next section. 

The environmental dependence of our color distributions is shown in fig. 3, 
where we compare with the SDSS data from  Balogh et al. (2004); for a given 
luminosity range, these have been binned according  to different environmental 
densities defined inside a given region around the fifth brightest galaxy. 

\vspace{-0.2cm} 
\begin{center} 
\scalebox{0.72}[0.72]{\rotatebox{-90}{\includegraphics{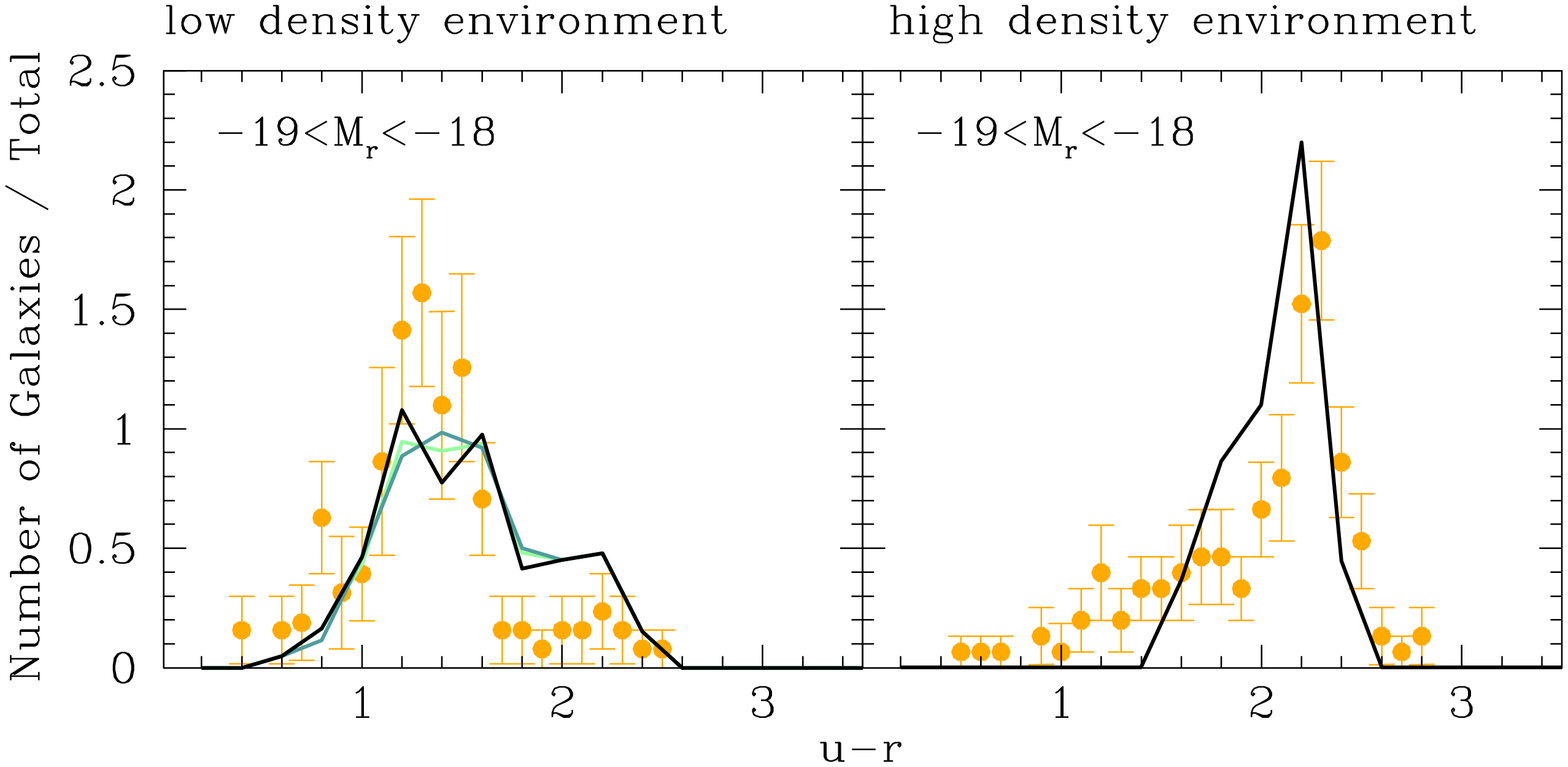}}} 
\end{center} 
{\footnotesize Fig.  3. - The dependence of the 
$u-r$ color distribution on the environment  is plotted for galaxies  in the bin 
$-18<M_r<-19$ to allow  a comparison with  the SDSS distribution (dots, from 
Balogh et al. 2004) in lower density (upper panel) and highest density (lower) 
environments where the  considered galaxies are  found (see text). The 
distributions have been normalized to yield unity when integrated over color. 
\vspace{0.2cm}}

Since our  SAM does  not include  any spatial information,  we cannot  perform a 
fully consistent  comparison, so  we proceed as follows. We focus on the less 
luminous bin $-18<M_r<-19$ considered by Balogh  et al.  (2004), since the 
environmental  effects produce the  most dramatic changes in  the color 
distribution at low luminosities. Then  we selet  the less massive and the most 
massive parent DM halos in our Monte Carlo run, and compare the color 
distributions of the galaxies contained in such environments with  the less and 
the most  dense environment  considered in  Balogh et  al. (2004). The highest 
density environment  considered by Balogh  et al. (2004)  corresponds to the 
core of rich clusters of galaxies, roughly consistent with our most  massive DM 
parent halo at $z=0$ which has a mass $1.8\cdot 10^{15}\,M_{\odot}$;  as for 
the  underdense regions,  we  select from  our  Monte Carlo  the  less dense 
environment (the less massive parent halo at $z=0$) containing a galaxy with 
magnitude in the considered range; the corresponding lower density environment 
in Balogh et al. (2004) has a projected surface density of 0.1 Mpc$^{-2}$.

Within  the  limits  of  our comparison,  the  model  correctly  reproduces the 
observed  shift   of  the   distribution  toward   redder  colors   when  denser 
environments are  considered. Indeed,  the environmental  dependence appearing 
in fig. 3 is stronger than the luminosity dependence; in fact, in the  densest 
environments (originated  from  the   most   biased  regions of  the  primordial 
field)  the distribution is  entirely  skewed toward red  ($u-r>2.3$) colors, 
even when low luminosities ($-18<M_r<-19$) are considered. A similar 
agreement holds for other luminosity ranges; in particular, the high-luminosity 
range ($M_r\lesssim -21$) is dominated by red galaxies in all environments. 

Finally,  we show  in fig.  4 the  model predictions  at higher  redshifts. At 
$z\approx 1$ we compare with data from Giallongo et al. (2005); their  composite 
sample (see  caption) covers  a range  spanning from  UV down  to the optical/IR 
wavelengths, with different depths  reaching $m_{H_{AB}}=26$ in the deepest 
field. The bimodal distribution computed from  the model persists  at $z\approx 
1$,  in agreement with  observations (see  also Bell  et al.  2004); indeed, 
our  model predicts the bimodality to be clearly present up to $z\approx 1.5$ 
(see dashed lines in the figure).

\vspace{0.1cm} 
\begin{center} 
\scalebox{0.49}[0.49]{\rotatebox{-90}{\includegraphics{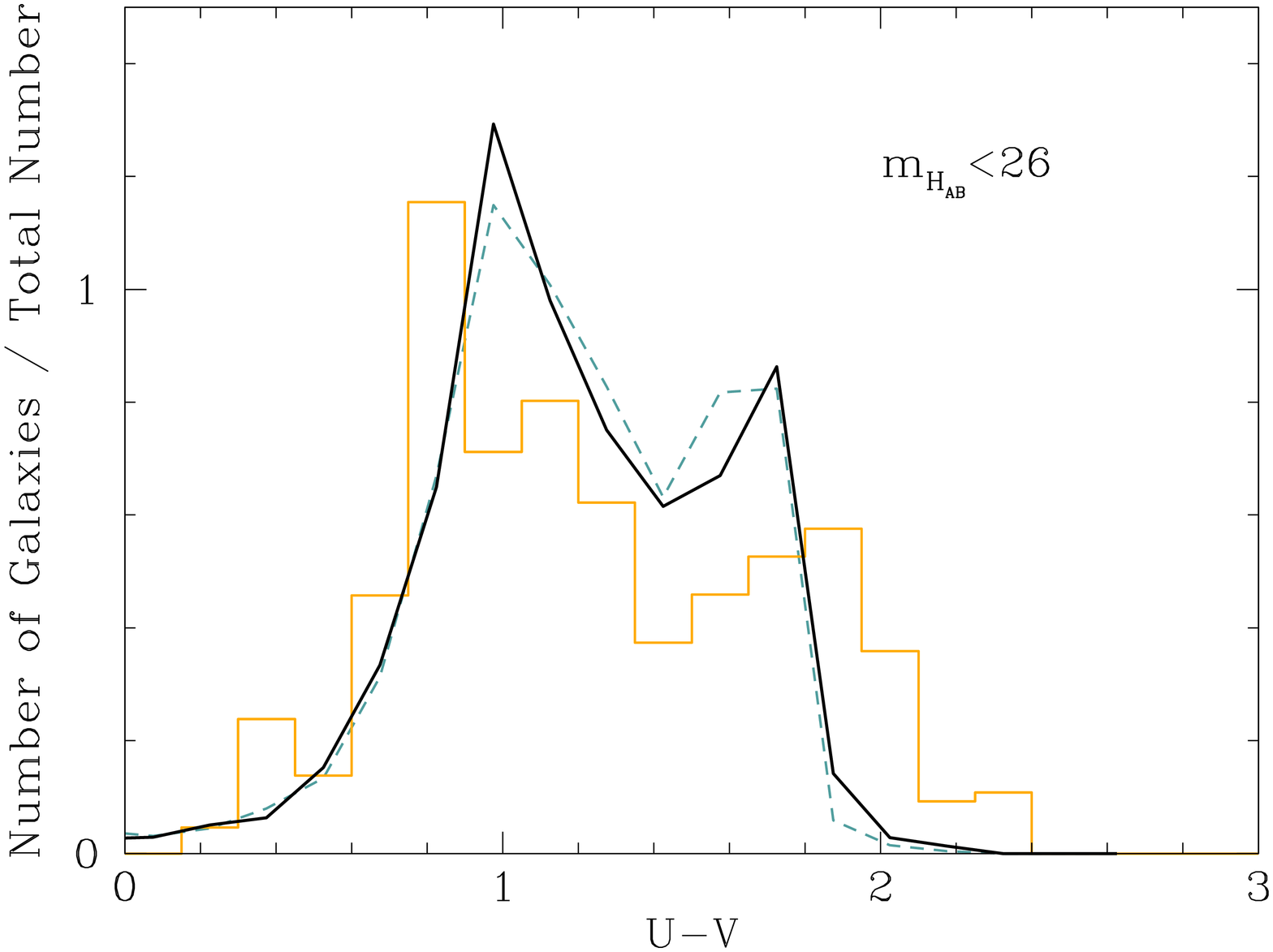}}} 
\end{center} 
{\footnotesize \vspace{-0.1cm} Fig.  4. - 
The $U-V$  color distribution at 
$0.8<z<1$  of $m_{H_{AB}}<26$ galaxies in  our model  is compared  with the 
data by  Giallongo et  al. (2005) derived from a composite  sample of galaxies 
including a  portion of the K20  field (Cimatti et al. 2002), and the HDFN, HDFS 
and Chandra fields; the Poisson error in each color bin was computed adopting 
the recipe by Geherels (1986) valid also for small numbers. We also show as a 
dashed curve the predicted color distribution at $z=1.5$ for the same magnitude 
limit.
\vspace{0.2cm}}

\subsection{Star Formation Histories of Blue and Red Galaxies}

The individual star--formation histories of the galaxies produced in our model 
provide a clear visualization of the evolutionary paths that led to the present-
day red and blue populations. We show in Fig. 5 a selection of individual star 
formation histories in our model, as obtained by summing over all the progenitor 
clumps which have merged to form galaxies of $M_r\simeq -20$ at $z=0$. As 
expected, the histories of the blue galaxies extend to much lower redshift than 
the red one, resulting in their bluer colors. We also remark the large variety 
of histories corresponding the blue population that is at the origin of the 
larger scatter observed in their color distributions at $z\approx 0$. On the 
other hand, the star formation histories of red galaxies are more similar 
(resulting in smaller scatter of the color function at $z\approx 0$, see fig. 
2), with large values of $\dot m_*$ at $z\gtrsim 2.5$; note that such histories 
are broadly peaked at $z\approx 4-5$ and sharply decline at $z\lesssim 2$, while 
the histories leading to blue local galaxies typically reach their maximum at 
$z\approx 2-3$, since at high-$z$ the star formation is effectively suppressed 
by Supernovae feedback (see Sect. 5). 

Note that bursts do not seem to constitute the physical origin of the bimodality 
in the color distribution at $z\approx 0$. Indeed, the red side of the 
population is mostly contributed by galaxies with smoothly evolving $\dot m_*$ 
(like that represented by the heavy dashed line in the bottom panel of fig. 
5), although several histories are characterized by major bursts, like the 
extreme star formation history (marked by the heavy solid line in the 
bottom panel) which has udergone a major bursts at $z\approx 2.5$. On the 
other hand, bursts seem to be crucial in providing extremely red objects (EROs) 
already at high $z\gtrsim 1.5$, as shown again by the history marked by the 
heavy solid line; see also the other history characterized by a major burst at 
$z\approx 2$ and by a drop of $\dot m_*$ at $z\gtrsim 1.3$. 

We checked that switching off starbursts triggered by encounters does 
not affect appreciably the bimodality of our color distribution at 
$z\approx 0$; this is consistent with our previous results (Menci et al. 
2004) which showed such events to alter mainly the properties of large-mass 
galaxies at $z\gtrsim 1$ without affecting the local distributions. The above 
conclusion is confirmed by the observation that galaxies residing in dense 
environments are characterized by red colors at $z=0$ even if their 
luminosity/mass is moderate (see fig. 4); such a property is difficult to 
explain only in terms of starbursts triggered by major mergings, since the 
merging probability of small-mass galaxies in high-density environments is 
strongly suppressed not only by their small cross section, but also by the large 
velocity dispersion charcteristic of galaxy clusters.

\vspace{0.2cm} 
\begin{center} 
\scalebox{0.56}[0.56]{\rotatebox{0}{\includegraphics{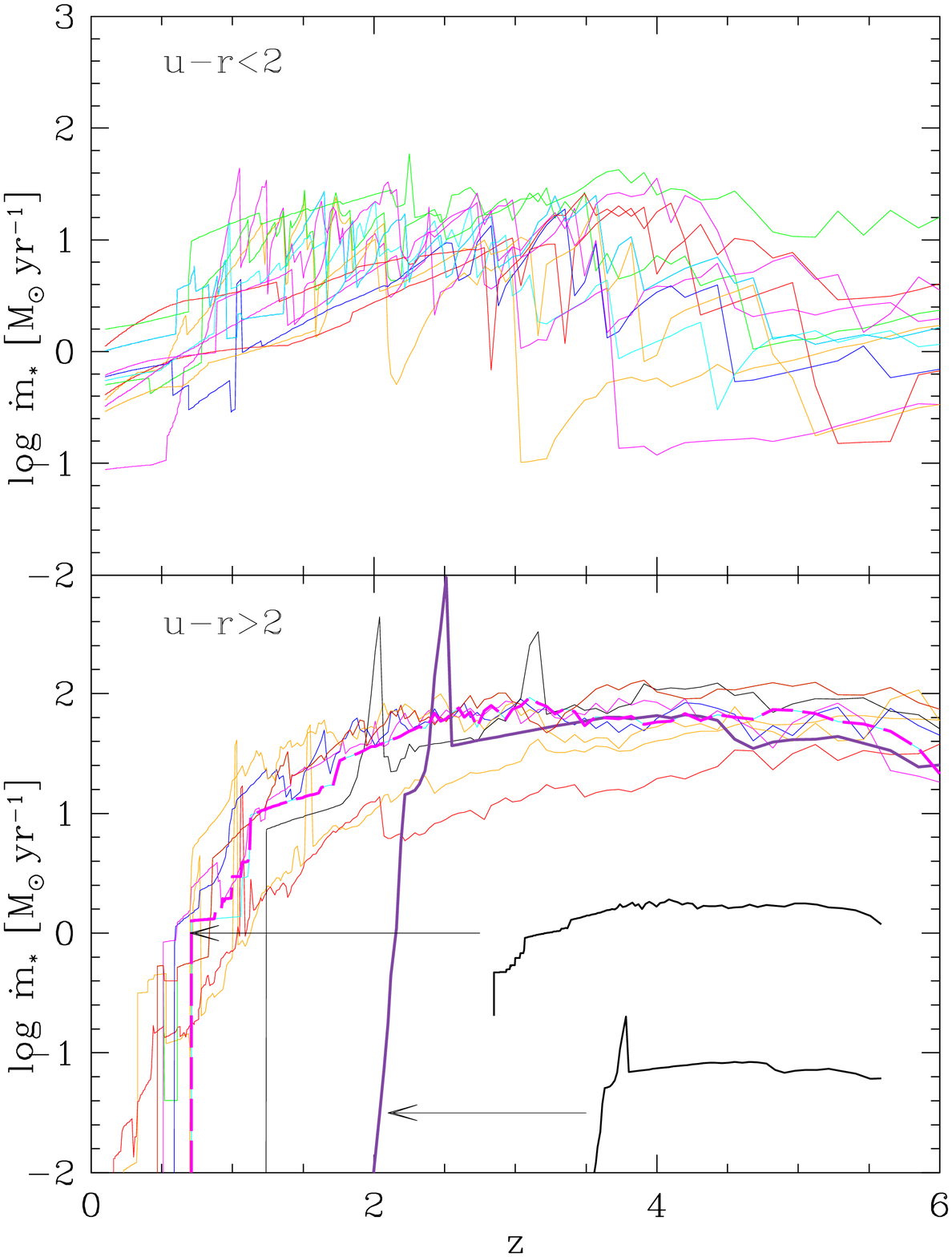}}} 
\end{center} 
{\footnotesize Fig. 5. - 
Star formation histories drawn 
from Monte Carlo realizations of galaxies which, at $z=0$, have luminosities 
in the range $-19.5\leq M_r\leq -20.5$. Each curve represents the sum of the 
$\dot m_*$ in all the progenitor clumps (at any given $z$) of galaxies belonging 
(at $z=0$) to the blue ($u-r<2$, top panel) and to the red ($u-r>2$, bottom 
panel) populations. Two star formation histories relevant to our 
discussion (see text) are highlightened by heavy curves (solid and dashed) and 
evidenced by the insets in the lower part of the plot.
\vspace{0.2cm}}

We now proceed to discuss the fundamental processes which are at the basis of 
the dichotomy between the star formation histories shown in the top and in the 
bottom panels of fig. 5, leading to the bimodal color distribution at low $z$. 

\section{HIERARCHICAL CLUSTERING AND THE ORIGIN OF BIMODALITY}

In hierarchical models, the scatter in the colors of galaxies with given final 
DM mass is due to the different realizations of the merging histories which lead 
to the considered mass; these ultimately trace back to the properties of the 
primordial density perturbations field where the galaxy progenitors first 
formed. 

In such a context, a correlation between the galaxy mass and the color is 
naturally predicted; in fact, the star formation in progenitor clumps later 
included into massive galaxies is peaked at higher redshift compared to that 
taking place in progenitors of small-mass galaxies, since the former form in 
the biased, high -density regions of the density field, collapsing at higher 
$z$ and containing denser gas. However, the absence of a typical mass scale 
in the progenitor distribution would not produce by itself a sharp transition 
in the star formation rate like that needed to originate the bimodal color 
distribution. 

A non-gravitational mass scale is naturally introduced in the star formation 
properties of different merging histories by the SNe feedback (as suggested by 
Dekel \& Silk 1986; see Dekel \& Birnboim 2005 and references therein). In low 
-mass galactic haloes such feedback rapidly reheats the cold gas not converted 
into stars, resulting into a self-regulated regime with $\Delta m_*\sim \Delta 
m_c$, where $\Delta m_c$ is the mass of gas cooled in the time step. The upper 
mass limit for such a regime to be effective can thus be estimated from 
$v_c^2\lesssim \epsilon_0\,\eta\,E_{SN}\approx 100$ km/s (see Sect. 1; for 
a more detailed derivation see Dekel \& Birnboim 2005), which, at high 
$z\approx 4-5$ corresponds to DM masses as low as $m_{0}\approx 
10^9\,M_{\odot}$. Thus in progenitor haloes with $m<m_0$ the star formation 
at high $z$ is self -regulated and is forced to remain below the limit 
set by the equilibrium between the feedback and the star formation (White \& 
Frenk 1991); on the other hand, in progenitors with $m>m_0$ the cold gas 
reservoir may grow (up to the maximal value $\Omega_b/\Omega$) at high $z$, 
and the star formation rate is limited only by the efficiency of cold gas 
conversion into stars (determined by the timescale $\tau_d$). 

Thus, hierarchical clustering naturally predicts: 1) the star formation 
in progenitor clumps later included into massive galaxies to peaked at 
higher redshift compared to that taking place in progenitors of small-mass 
galaxies; and 2) the existence of a mass threshold $m_0$ below which the cold 
gas content of the DM clumps at high $z$ is effectively reheated an the star 
formation is self-regulated. Such generic features are in qualitative agreement 
with the observation that bright galaxies are typically redder (de 
Vaucouleurs 1961; Bower, 1992).

In our model, the above two points combine with 3) a star formation 
efficiency that enforces the partition between gas rich and gas poor systems, 
above and below the threshold $m_0$. Spcifically, we adopt $\tau_d\propto m/m_c$ 
derived from the disk model developed by Mo, Mao \& White (1998) (as described 
in Sect. 2).

\vspace{0.2cm} 
\begin{center} 
\scalebox{0.62}[0.62]{\rotatebox{-90}{\includegraphics{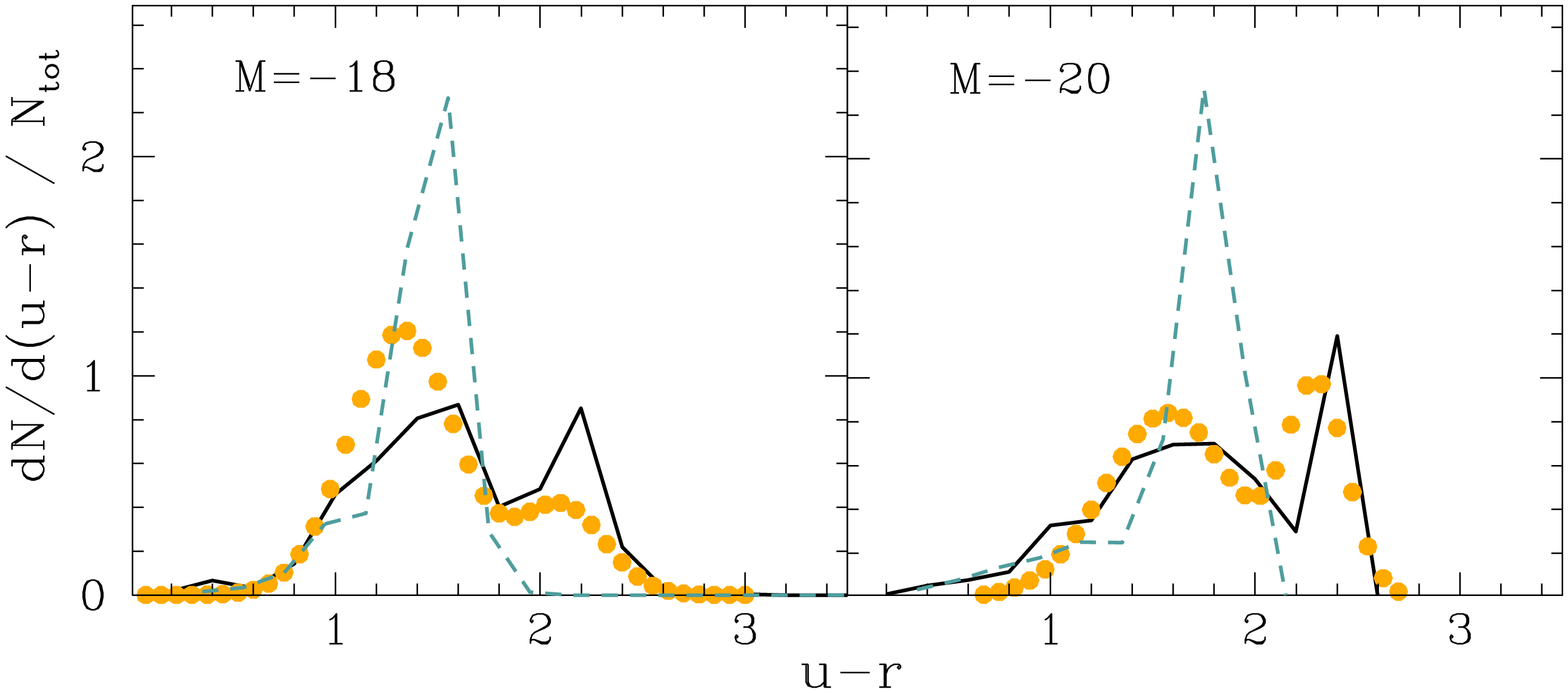}}} 
\end{center} 
{\footnotesize Fig. 6. - 
The color distribution of our 
fiducial model (solid line) -- i.e., star formation timescale $\tau_d\propto 
m/m_c$ derived from the disk model developed by Mo, Mao \& White (1998) -- is 
compared with that obtained by the model on assuming a constant ratio 
$m_c/m=0.05$ (dashed lines). The shaded area represents the uncertainty in the 
Gaussian fits to the SDSS data (Baldry et al. 2004). 
\vspace{0.2cm}}

In fact, at high-$z$, the cold gas reservoir of progenitors clumps with $m<m_0$ 
is continuously depleted by effective feedback, as discussed above; this 
increases the timescale $\tau_d\propto m/m_c$, thus suppressing star formation 
and avoiding the sudden conversion of cold gas into stars. At later cosmic 
times, as the progenitor masses grow above the threshold $m_0$, cold gas 
begins to accumulate in the progenitor clumps and is made available for star 
formation, which continues down to low redshifts. Such smooth star formation 
histories (the upper panel of fig. 5) lead to the local faint blue population. 

On the other hand, in progenitor clumps with $m>m_0$ at high-$z$ the cold gas is 
not effectively reheated; thus, the rapid cooling taking place at high-$z$ leads 
to large $m_c/m$ ratios, which shorten the star formation timescale. The cold 
gas is rapidly converted into stars, and begins to exhaust at $z\lesssim 2$ 
(see fig. 5); at later times the star formation is further suppressed since 
now such galaxies have low $m_c/m$ ratios. Thereafter, such galaxies undergo an 
almost quiescent phase characterized by a fast drop of $\dot m_*$; such 
histories are typical of massive galaxies and of galaxies forming in biased 
regions of the primordial density field (which later become the galaxy 
environment), and originate the red population at $z\approx 0$. 

The more effective is the star formation in high-$z$, gas rich haloes, the 
sharper is the transition between the feedback-regulated regime and the 
supply-limited regime. In fact, when we adopt a star formation timescale not 
explicitly depending on $m_c/m$ (like in most current SAMs), we obtain a smooth 
(non-bimodal) color distribution; this is shown in fig. 6, where the color 
distribution from our model is compared with that obtained on adopting a fixed 
value for the ratio $m_c/m=0.05$ (the average value in Mo, Mao \& White 1998) in 
the expression for the function $g(v)$ defined in Sect. 2. Note that this model 
still predictes a correlation between the color and the luminosity of present-day 
galaxies: as we discussed above, this is indeed a generic feature of 
hierarchical scenarios. In our fiducial model, we achieve redder color in an 
appreciable fraction of galaxies compared to the model with fixed $m_c/m$. 

Summarizing, in our model the star formation in massive haloes at low 
redshift is suppressed by three concurring processes: i) at high -$z$, the 
rapid conversion of gas into stars, enhanced by our star formation timescale 
$\tau_d\propto m/m_c$, leads to a fast exhaustion of the cold gas 
reservoir; ii) at $z\lesssim 2$, the cold gas reservoir of massive galaxies 
is depleted by major merging events (see Sect.2); iii) as a consequence of 
i) and ii), the ratio $m_c/m$ in such galaxies is particularly low, 
leading to a longer star formation timescale $\tau_d\propto m/m_c$, which 
further supresses star formation. 

It must be emphasized that results similar to our model might be obtained by 
any mechanism enforcing the dependence of $\dot m_*$ on the cold gas 
accretion (like the cold streams discussed by Dekel \& Birnboim 2005). 

Thus, the origin of bimodality is connected to the interplay between the 
properties of the merging histories involving the progenitors of local 
galaxies and the feedback/star formation process. To further test such a 
picture we proceed as follows: we focus on the merging histories of 
galaxies with luminosity (at $z=0$) in the bin centered at $M_r=-20$ (so that 
the considered galaxies have similar total mass $m\approx 10^{12}\,M_{\odot}$), 
and we select the galaxies belonging to the blue and to the red local 
population. We then compute the mass distribution of the progenitors of 
the (locally) red and blue galaxies, separately, which are shown at 
different redshifts in fig. 7. 

\vspace{1.cm} 
\begin{center} 
\scalebox{0.57}[0.57]{\rotatebox{0}{\includegraphics{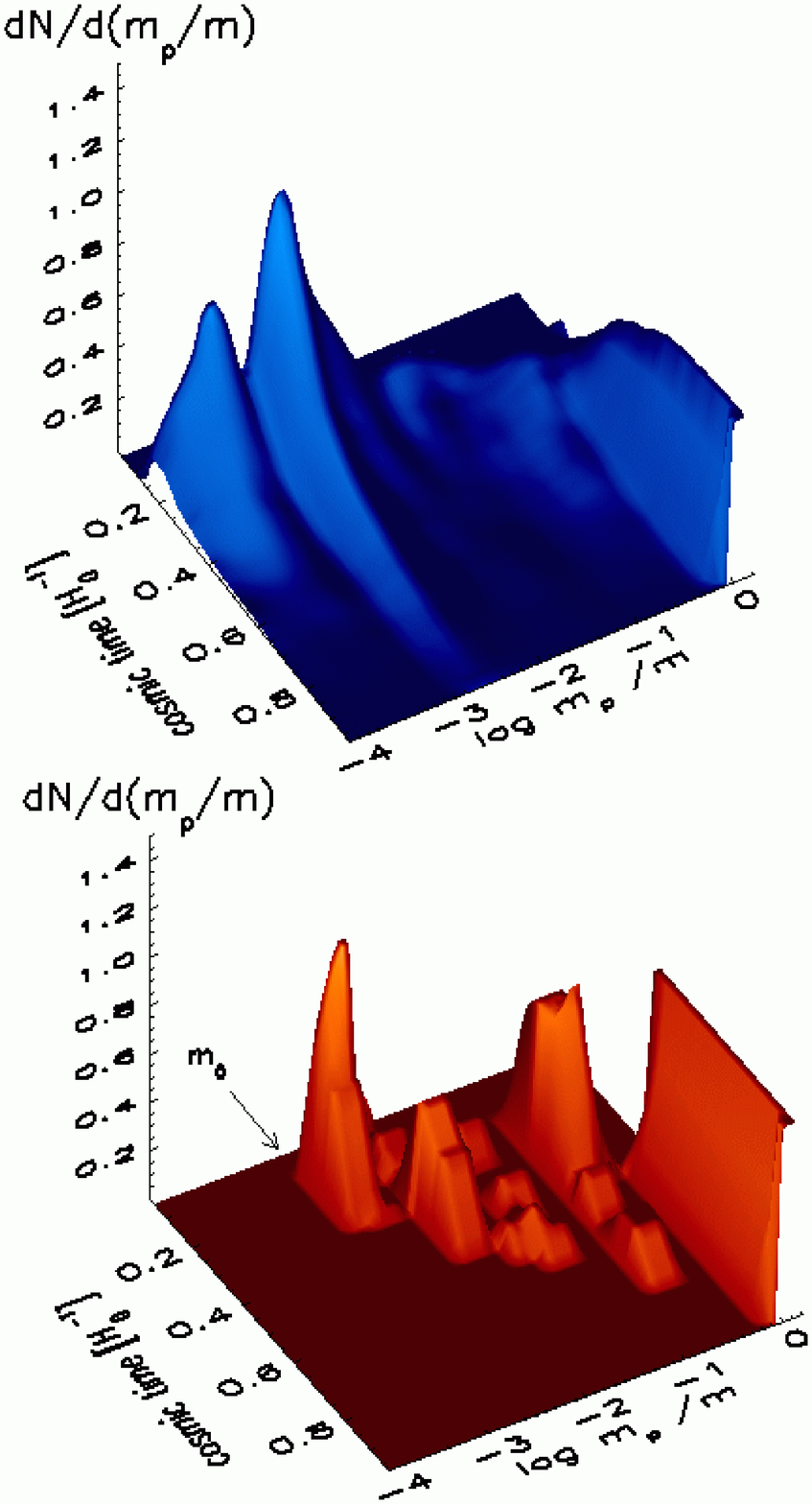}}} 
\end{center} 
{\footnotesize Fig. 7. - 
The distribution $dN/d(m_p/m)$ 
of progenitors of local galaxies with luminosity $-19.75<M_r<-20.25$ as a 
function of the progenitor mass $m_p$ (normalized to the final mass of the 
galaxy $m$) and of the cosmic time. Top panel refers to local galaxies with blue 
($u-r<2$) color, bottom panel to the red ones ($u-r\geq 2$). Time is in units of 
$1/H_0$; in such units, $t=0.1$ corresponds to $z=4.2$, and $t=0.4$ to $z=1$. 
\vspace{0.2cm}}

As expected in our picture, the two classes of merging histories are 
qualitatively different. At the highest redshifts the mass distribution of the 
progenitors of the (present--day) red galaxies is skewed toward larger masses 
compared with the the blue ones, and they rapidly reach their final mass. 
Conversely, the merging histories of present--day blue galaxies are 
characterized by the infall of low-mass progenitors well below $z\approx 1$, 
leading to a refueling of gas over a longer cosmic time. Moreover, the 
existence of a low-mass threshold for the progenitors of the red objects 
(corresponding to the mass scale $m_0$) is clearly shown by the plot; the 
numerical value of $m_0$ obtained from the bottom panel in fig. 7 corresponds to 
the critical velocity of 100 km/s derived above by requiring the reheating from 
feedback to regulate the star formation rate.

This confirms our picture, and enlightens that the low-$z$ colors of galaxies 
are the final result of a partition in the star formation histories 
of the progenitors occurring at high $z\approx 4-5$. Such a partition is 
originated by the properties of the DM density field, the existence of a 
mass threshold for self-regulating star formation, and a star formation 
efficiency that enforces the partition between gas rich and gas poor systems, 
above and below the threshold $m_0$.

\section{SUMMARY AND CONCLUSIONS}

We quantitatively showed how the observed properties of the bimodal 
color distribution of galaxies arises, in the framework of the 
hierarchical clustering picture, from the interplay between the 
properties of the merging histories involving the progenitors of local 
galaxies and the feedback/star formation process. 

Our SAM successfully matches several fundamental properties of the galaxy 
color distribution: its bimodal shape at $z\approx 0$ defining two classes of 
objects whose partition occurs at $u-r\approx 2$, with a small spread in the 
distribution of red objects compared to the blue ones (fig. 2); the luminosity 
dependence of the bimodal distribution, characterized by the red population 
growing with increasing $L$ and dominating the distribution for 
$M_r\lesssim -21$ (fig. 2); its environmental dependence (characterized by 
redder galaxies in denser environments) which overrides the luminosity 
dependence in the sense that even the color distribution of local faint 
galaxies ($-18<M_r<-19$) is skewed toward red ($u-r\approx 2.2$) colors (fig. 
3) when only dense environments are considered; the persistence of the bimodal 
distribution up to $z\approx 1.5$ (fig. 4). As we showed in a previous paper, 
the abundances of the red and blue galaxies are consistent with the observations 
up to $z\approx 3$ (Giallongo et al. 2005). 

We have also shown that the two color populations result from two classes 
of star formation (fig. 5 ) and merging histories:  
\newline 
i) Merging 
histories leading to the local red population are characterized by a progenitor 
distribution skewed toward larger masses at high $z\approx 4-5$ (fig. 7, 
bottom). In such progenitors, feedback is ineffective in rehating/expelling 
the cold gas content and, if the latter is rapidly converted into stars at 
$z\approx 4-5$, the main progenitors exhaust their cold gas reservoir at 
$z\approx 2$ (see fig. 5) and thereafter undergo an almost quiescent 
phase characterized by a fast drop of $\dot m_*$. Massive galaxies and 
galaxies formed in a biased region of the primordial density field 
(which later become the galaxy environment) are preferentially 
assembled through the former kind of merging history. This straightforwardly 
explains the luminosity and environmental dependence of the color 
distribution. \newline ii) Merging histories leading to the local blue 
population are characterized by a progenitor distribution dominated by small-
mass progenitors (fig. 7, top). There the star formation is self-regulated by 
feedback, which limits the cold gas content and is effective in prolonging the 
active phase of star formation. In addition, such merging histories are 
characterized by the infall of low-mass progenitors down to $z\approx 1$, 
resulting into a refueling of gas extended over a longer cosmic time. The 
corresponding star formation histories are illustrated in the top panel of fig. 
5. Small-mass galaxies are generally built-up through this evolutionary 
path. 

Hierarchical clustering provides a framework for the rising of bimodality 
because of two natural features: 1) star formation histories of massive galaxies 
and of galaxies formed in biased, high density regions are peaked (on average) 
at higher $z$ compared to lower mass galaxies; 2) the existence of a non-
gravitational mass scale $m_0$ (whose value corresponding to $v\approx 100$ km/s 
is set by the Supernovae feedback) such that for $m<m_0$ the star formation is 
self-regulated and the cold gas content is continuously depleted by effective 
feedback. These features are generic to most hierarchical clustering models  and 
typically contribute to yield to a global color-magnitude relation where 
brighter galaxies tend to be redder.

Within such a framework, star formation laws which enhance the gas consumption 
effects described above are necessary to yield a bimodality in the color 
distribution of present-day galaxies. In particular, we have shown that 
enforcing the dependence of $\dot m_*$ on the cold gas accretion above 
the scaling $\dot m_*\propto m_c$ lead to a sharper partition in the star 
formation histories of clumps with progenitor masses above and below $m_0$. 
Thus, the bimodal color distribution of galaxies originates at high redshifts 
$z\approx 4-5$ during the formation and the merging of their progenitors. 
This naturally explains the observed appearence of a bimodal distribution 
already at $z\approx 1.5$. 

\acknowledgments We thank A. Cavaliere and R. Somerville for useful discussions. 
We acknowledge grants from MURST.


\newpage

\begin{references} 
\reference{}Baldry, I.K., Glazebrook, K., Brinkmann, J., Zeljko, I., Lupton, 
R.H., Nichol, R.C., Szalay, A.S. 2004, ApJ, 600, 681 
\reference{}Balogh, M.L., Baldry, I.K., Nichol, R.C., Miller, C., 
Bower, R., Glazebrook, K. 2004, ApJ, 615, L101 
\reference{}Bell, E., Wolf, C., Meisenheimer, K., Rix, H.-W., Borch, A., Dye, 
S., Kleineinrich, M., McIntosh, D. 2004, ApJ, 608, 752 
\reference{}Benson, A.J., Bower, R.G., Frenk, C.S., Lacey, C.G., Baugh, C.M., 
Cole, S. 2003, ApJ, 599, 38
\reference{}Blanton, M.R. et al. 2001, ApJ, 121, 2358 
\reference{}Bond, J.R., Cole, S., Efstathiou, G., \& Kaiser, N., 1991, ApJ, 379, 
440 \reference{}Bower, R.J., Lucey, J.R., Ellis, R.S. 1992, MNRAS, 254, 601 
\reference{}Bruzual, A.G., \& Charlot, S., 1993, ApJ, 105, 538 
\reference{}Calzetti, D., 1997, in Proc. of "The Ultraviolet Universe at Low 
and High Redshift : Probing the Progress of Galaxy Evolution", AIP Conference 
Proceedings, 408, p.403 
\reference{} Cimatti, A. et al. 2002, \aap, 392, 395 
\reference{}Cole, S., Aragon-Salamanca, A., Frenk, C.S., Navarro, J.F., Zepf, 
S.E., 1994, MNRAS, 271, 781 
\reference{}Cole, S., Lacey, C.G., Baugh, C.M., Frenk, C.S., 2000, MNRAS, 319, 168 
\reference{}Cole, S. et al. 2001, MNRAS, 326, 255 
\reference{}Cox, T.J., Primack, J., Jonsson, P., Somerville, R.S. 2004, ApJ, 
607, L87 
\reference{}de Vaucouleurs, G. 1961, ApJS, 5, 223 
\reference{}Dekel, A., \& Birnboim, Y. 2005, ApJ, in press [astro-ph/0412300] 
\reference{}Gehrels, N. 1986, ApJ, 303, 336 
\reference{}Giallongo, E., Salimbeni, S., Menci, N., Zamorani, G., Fontana, A., Dickinson, M., 
Cristiani, S., Pozzetti, L. 2005, ApJ, 622, 116 
\reference{}Giallongo, E., Menci, N., Poli, F., D'Odorico, S., Fontana, A. 2000, 
ApJ, 530, L73 
\reference{}Giovanelli, R., Haynes, M. P., da Costa, L. N., 
Freudling, W., Salzer, J. J., \& Wegner G., 1997, AJ, 113, 22 
\reference{}Kauffmann, G., White, S.D.M., \& Guiderdoni, B., 1993, MNRAS, 264, 
201 \reference{}Kauffmann, G. 1996, MNRAS, 281, 475 
\reference{}Kauffmann, G., Charlot, S. 1998, MNRAS, 294, 705 
\reference{}Lacey, C., \& Cole, S., 1993, MNRAS, 262, 627 
\reference{}Madgwick, D.S., et al., 2002, MNRAS, 333, 133
\reference{}Mathewson, D.S., Ford, V.L., \& Buchhorn, 1992, ApJS, 81, 413 
\reference{}Menci, N., Cavaliere, A., Fontana, A., Giallongo, E., Poli, F., 
2002, ApJ, 578, 18 
\reference{}Menci, N., Cavaliere, A., Fontana, A., Giallongo, E., Poli, F., 
Vittorini, V. 2004, ApJ, 604, 12 
\reference{}Mo, H.J, Mao S., \& White, S.D.M., 1998, MNRAS, 295, 319 
\reference{}Nagamine, K., Springel, V., Hernquist, L., Machacek, M. 2004, MNRAS, 
350, 385 
\reference{}Nagamine, K., Cen, R., Hernquist, L., Ostriker, J.P., 
Springel, V., 2005, ApJ, submitted [astro-ph/0502001] 
\reference{}Navarro, J.F., Frenk, C.S., \& White, S.D.M. 1997, ApJ, 490, 493 
\reference{}Okamoto, T., Nagashima, M. 2003, ApJ, 587, 500 
\reference{}Press, W.H., \& Schechter, P., 1974, \apj, 187, 425 
\reference{}Schneider,  S.E., 1997, PASA, 14, 99 
\reference{}Somerville, R.S., \& Primack, J.R., 1999, MNRAS, 310, 1087 
\reference{}Somerville, R.S., Primack, J.R., \& Faber, S.M., 2001, MNRAS, 320, 
504 
\reference{}Strateva, I. et al. 2001, AJ, 122, 1861 
\reference{}Steidel, C.C., Adelberger, K.L., Giavalisco, M., 
Dickinson, M., Pettini, M., 1999, ApJ, 519, 1 
\reference{}White, S.D.M.,  \& Frenk, C.S. 1991, ApJ, 379, 52 
\reference{}Willick, J.A., Courteau, S., Faber, S.M., Burstein, D., Dekel, A., 
\& Kolatt, T., 1996, ApJ, 457, 460 
\reference{}Zwaan, M.A., Briggs, F.H., Sprayberry, D., Sorar, E. 1997, ApJ, 490, 173 
\end{references}
\end{document}